\documentclass[11pt]{article}
\usepackage{amssymb} 
\def\R{\mbox{$\mathbb R$}}

\newtheorem{claim}{Claim}

\begin{document}

\title{The union of unit balls has quadratic complexity, even if they all contain the origin}
\date{June, 1999}
\author{Hervé Br{\"o}nnimann and Olivier Devillers\thanks{
INRIA, BP 93, 06902 Sophia Antipolis, France.
\texttt{\{Herve.Bronnimann|Olivier.Devillers\}@sophia.inria.fr}}}
\maketitle


\begin{abstract}
We provide a lower bound construction showing that the union
of unit balls in $\R^3$ has quadratic complexity, even if they all contain
the origin. This settles a conjecture of Sharir.
\end{abstract}

\section{Introduction}

The union of a set of $n$ balls in $\R^3$ has quadratic complexity
$\Theta(n^2)$, even if they all have the same radius. All the already known
constructions have balls scattered around, however, and Sharir posed the
problem whether a quadratic complexity could be achieved if all the balls (of
same radius) contained the origin.

In this note, we show a construction of $n$ unit balls, all containing the
origin, whose union has complexity $\Theta(n^2)$.  As a trivial observation, we
observe that the centers are arbitrarily close to the origin in our
construction. In fact, if the centers are forced to be at least pairwise
$\varepsilon$ apart, for some constant $\varepsilon>0$, then no more than $O(\frac 1{\varepsilon^3})$ can meet in
a single point, and hence the union has complexity at most $O(\frac
1{\varepsilon^3} n)=O_\varepsilon(n)$.  It is an interesting open question what
a condition should be so that the union have subquadratic complexity and yet the
balls have arbitrarily close centers.

By contrast, the {\em intersection} of $n$ balls can have quadratic complexity
if their radii are not constrained, but the complexity is linear if all the
radii are the same \cite{h-bevv-56}. Similarly, the convex hull of $n$ balls
can have also quadratic complexity \cite{bcddy-acchs-96}, but that complexity is
linear if they all have the same radius.

\section{Construction}

Let $m$ and $k$ be any integers.
We define two families of unit balls: the first consists of $k$ unit balls
whose centers lie on a small vertical segment; the second
consists of $m$ unit balls whose centers lie on a small circle under
the segment. (See Figure \ref{fig-final}.)
We show below that their union has quadratic $O(km)$
complexity.

\begin{figure}[t]
\centerline{\def\IPEfile{union1k.ipe}\input{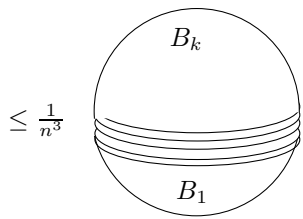}}
\caption{\label{fig-vertical}The union $\cup_{1\leq i\leq k} B_i$.}
\end{figure}

\paragraph{The balls $B_1$ \ldots $B_{k}$.}
We denote by $B(p,r)$ the ball centered at $p$ and of radius $r$.
Let $n=k+m$ and
$P_i$ denote the point of coordinates $(0,0,(i-1)/n^4)$, and
$B_i=B(P_i,1)$, for $i=1,\ldots,k$. It is clear that the boundary
of $\cup_{1\leq i\leq k} B_i$ consists of two hemispheres
belonging to $B_1$ and $B_k$ linked
by a narrow cylinder of height less than $k/n^4 \leq 1/n^3$. This cylinder
contains all the circles $\partial B_i\cap \partial B_{i+1}$ for $i=1,\ldots,k-1$.
(See Figure \ref{fig-vertical}.)

\paragraph{The balls $B_{k+1}$ \ldots $B_{k+m}$.}
Let $R$ be the point of coordinates $(x,0,z)$ with
$$
   x = \frac{2n^2-4}{n^4}, \qquad z = -\frac{2n^2-4}{n^3} \, .
$$
(Any values satisfying the constraints $P_kH<1$ in (\ref{eq:claim1}) and $\ell<\frac2n$
in (\ref{eq:claim2}) below would do.)
We define $\theta$ as the rotation around the $z$-axis of angle $2\pi/m$, and
for each $j=1,\ldots,m$, $R_{k+j}=\theta^{j-1}(R)$ and $B_{k+j}=B(R_{k+j},1)$.

\section{Analysis}

\begin{figure}[t]
\centerline{\def\IPEfile{sector.ipe}\input{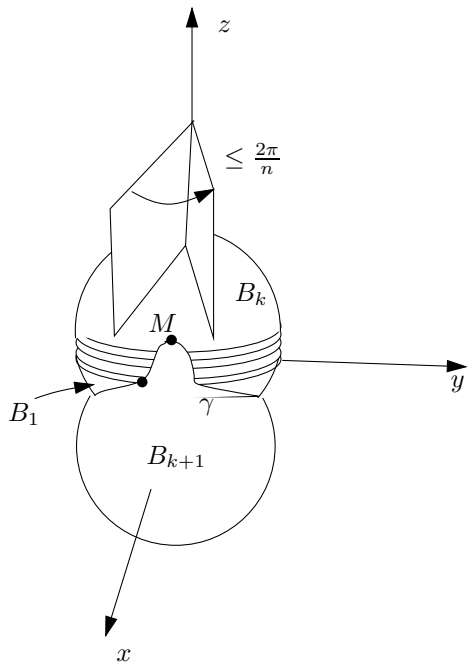}}
\caption{\label{fig-sector}The union $\bigcup_{1\leq i\leq k} B_i \cup B_{k+1}$.
The curve $\gamma$ consists of a
portion that belongs to $B_1$ and of another portion which is
contained in a dihedral sector of angle less than $\pi/m$.}
\end{figure}

By our choice of $x$ and $z$, we prove below that the boundaries of $B_{k+1}$
and of the union $\cup_{i=1}^k B_i$ depicted in Figure \ref{fig-vertical} meet
along a curve $\gamma$ which satisfies the two claims below. The situation is 
depicted on Figure \ref{fig-sector}.

\begin{claim}
The curve $\gamma$ intersects all the balls $B_i$ for $i=0,\ldots,k$, .
\label{claim:1}
\end{claim}

\begin{claim}
The portion of $\gamma$ which does not belong to $B_1$ (equivalently,
which belongs to the union $\cup_{i=2}^k B_i$) is contained in an angular sector
of angle at most $2\pi/m$.
\label{claim:2}
\end{claim}

\begin{figure}[t]
\centerline{\def\IPEfile{union.ipe}\input{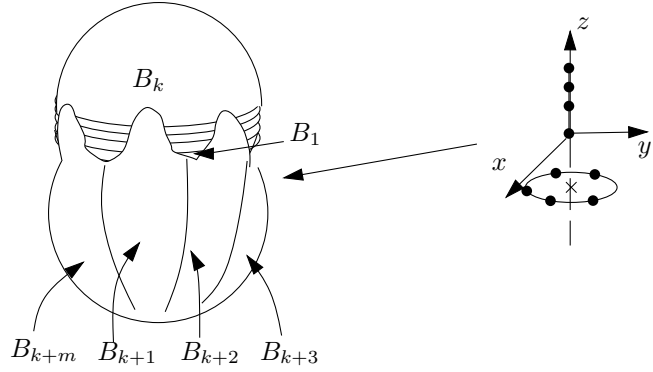}}
\caption{\label{fig-final}The union $\cup_{1\leq i\leq k+m} B_i$. On the right, a
blow-up of the centers.}
\end{figure}

From claim \ref{claim:2}, we conclude that the portion of $\gamma$ which does
not belong to $B_1$ is contained in the boundary of the union of the $n=k+m$
balls, 
From claim \ref{claim:1}, we conclude that  the portion of $\gamma$ which does
not belong to $B_1$ has complexity $\Omega(k)$. From claim \ref{claim:2}, that
it is contained in a small angular sector, hence appears completely on the boundary of the union of the $n=k+m$ balls, and it is replicated $m$ times, once for each
of the balls $B_j$, $j=1,\ldots,m$. It follows that the union of all the balls
$B_i$ for $i=1,\ldots,k+m$ has quadratic complexity $\Omega(km)$. Moreover, all the
balls contain the origin.
The union of the $n$ balls is depicted on Figure \ref{fig-final}.

\medskip

\begin{figure}[t]
\centerline{\def\IPEfile{claim1.ipe}\input{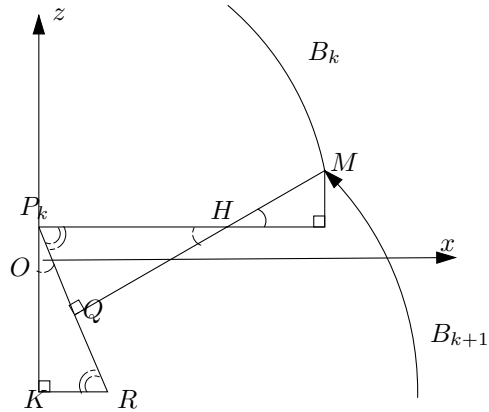}}
\caption{\label{fig-claim1}Figure for Claim \ref{claim:1}.}
\end{figure}

\begin{figure}[t]
\centerline{\def\IPEfile{claim2.ipe}\input{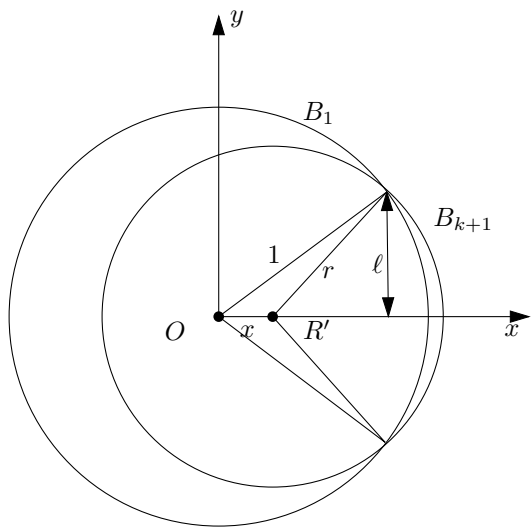}}
\caption{\label{fig-claim2}Figure for Claim \ref{claim:2}.}
\end{figure}

The proofs involve only elementary geometry and trigonometry. The situation is
depicted in Figure \ref{fig-claim1} and \ref{fig-claim2}. Figure \ref{fig-claim1}
depicts a section in the $xz$-plane of the spheres $\partial B_i$ and $\partial B_{k+1}$ and
the point $M$, the highest point of intersection of the bounding
spheres. The point $M$ is also depicted on Figure \ref{fig-sector}.

\paragraph{Proof of Claim \ref{claim:1}.} 
It suffices to prove that $M$ is higher than $P_k$, since then $\gamma$ extends
higher than $P_k$ as well and passes through $M$ by symmetry. The lowest point
of $\gamma$ belongs to $B_1$ and is clearly below the origin. The two facts
together prove that $\gamma$ must intersect all the balls between $B_1$ and
$B_k$.

Let $H$ be the point in the $xz$-plane on the median bisector of $R$ and $P_k$,
with same $z$-ordinate as $P_k$. (See Figure \ref{fig-claim1}.) In order to prove that $M$ is higher than
$P_k$, it suffices to prove that $H$ belongs to $B_{k}$, since then $M$ is
farther along the bisector. The two triangles $QP_kH$ and $KRP_k$ have equal
angles, hence they are similar. It follows that
\begin{equation}
    P_kH = P_kR \frac{P_kQ}{RK} = \frac{P_kR^2}{2RK}
    =\frac{x^2+\left(z-z_k\right)^2}{2x},
\label{eq:claim1}
\end{equation}
where $z_k=\frac{k-1}{n^4}$.
For $x$ and $z$ as given in the construction, we have
\[
   P_kH = 1/16\,{\frac {-40\,{n}^{4}-15\,{n}^{2}+68+16\,{n}^{6}-16\,{n}^{3}+28\,n}{{n}^{4}\left ({n}^{2}-2\right )}}
\]
which is smaller than 1 for $n\geq 2$.

\paragraph{Proof of Claim \ref{claim:2}.} 
It is easy to see that the intersection of $\gamma$ and a ball $B_i$ ($2\leq
i\leq k$) consists of at most two arcs of circle, any of which is monotone in
angular coordinates around the $z$-axis, and that any such arc is entirely above
the plane $z=0$. Hence the intersections of $\gamma$ with the
$xy$-plane belong to $B_1$ and $B_{k+1}$. It suffices to show that these intersections
are at a distance $\ell$ at most $\frac2n\leq \sin \frac\pi{m}$ from
the $x$-axis. (See Figure \ref{fig-claim2}.)

In the $xy$-plane section, $B_1$ is a unit circle, and $B_{k+1}$ is a circle of
radius $r=\sqrt{1-z^2}$ and center $R'$ of coordinates $(x,0)$. (Recall that the center of $B_{k+1}$
has coordinates $(x,0,z)$.) Hence $\ell$ is the height of a triangle with base
$x$ and sides 1 and $r<1$.  It is elementary to compute that
\begin{equation}
    \ell = \sqrt{1 - \left( \frac{z^2+x^2}{2x} \right)^2 } \, .
    \label{eq:claim2}
\end{equation}
For our choice or $x$ and $z$, this yields
\[
  \ell = \sqrt {{\frac {2\,{n}^{6}+3\,{n}^{4}-4\,{n}^{2}-4}{{n}^{8}}}}
\]
which is smaller than $2/n$ for $n\geq 2$.

\bigskip

\paragraph{Acknowledgments.} Thanks to Micha Sharir for pointing
out the problem the us. It was also pointed out that Alon Efrat
might have a construction which leads to a quadratic lower bound as
well. We have derived our construction independently.

\bibliographystyle{plain}
\bibliography{geom}

\end{document}